\documentclass[showpacs,twocolumn,aps,prl]{revtex4}
\usepackage{mathrsfs}
\usepackage{amsmath}
\usepackage{amsfonts}
\usepackage{amssymb}
\usepackage{amsthm}
\usepackage{graphicx}
\usepackage{color}
\usepackage{epsfig}
\usepackage{cancel}
\usepackage{bm}
\providecommand{\U}[1]{\protect\rule{.1in}{.1in}}
\usepackage{ulem}
\begin{document}
\title{A Generic Phase between Disordered Weyl Semimetal and Diffusive Metal} 
\author{Ying Su$^{1,2}$}
\author{X. S. Wang$^{3,1}$}
\author{X. R. Wang$^{1,2}$}
\email[corresponding author: ]{phxwan@ust.hk}
\affiliation{$^{1}$Physics Department, The Hong Kong University of
Science and Technology, Clear Water Bay, Kowloon, Hong Kong}
\affiliation{$^{2}$HKUST Shenzhen Research Institute, Shenzhen 518057,
China}
\affiliation{$^3$School of Microelectronics and Solid-State Electronics,
University of Electronic Science and Technology of China, Chengdu,
Sichuan 610054, China}
\date{\today}
\begin{abstract}
Quantum phase transitions of three-dimensional (3D) Weyl semimetals 
(WSMs) subject to uncorrelated on-site disorder are investigated 
through quantum conductance calculations and finite-size scaling of 
localization length. Contrary to previous claims that a direct 
transition from a WSM to a diffusive metal (DM) occurs, an 
intermediate phase of Chern insulator (CI) between the two distinct 
metallic phases should exist due to internode scattering that is 
comparable to intranode scattering. The critical exponent of  
localization length is $\nu\simeq 1.3$ for both the WSM-CI and 
CI-DM transitions, in the same universality class of 3D 
Gaussian unitary ensemble of the Anderson localization transition. 
The CI phase is confirmed by quantized nonzero Hall conductances 
in the bulk insulating
phase established by localization length calculations. The disorder-induced various 
plateau-plateau transitions in both the WSM and CI phases are 
observed and explained by the self-consistent Born approximation. 
Furthermore, we clarify that the occurrence of zero density of 
states at Weyl nodes is not a good criterion for the disordered WSM, 
and there is no fundamental principle to support the hypothesis 
of divergence of localization length at the WSM-DM transition. 
\end{abstract}

\vspace{-0.2in}
\maketitle
Weyl semimetals (WSMs), characterized by the linear crossings of their 
conduction and valence bands at Weyl nodes (WNs) and the inevitable 
generation of topologically protected surface states, have attracted 
enormous attention in recent years because of their exotic properties 
and possible applications \cite{W1,W2,W3,W4,W5,W5.5,W6,W7,W8,W9,W10,W11}. 
Interestingly, WSM crystals are quite common instead of rare. 
The reason is that the most generic Hamiltonian describing two bands 
of a crystal is the direct sum of $2\times 2$ matrices in the momentum 
space as $H=\cup_{\bm{k}}\oplus h(\bm{k})$, where $\bm{k}$ is the 
lattice momentum. Thus, $h(\bm{k})$ must take a form of 
$\varepsilon_0(\bm{k})I+ \sum_\alpha h_\alpha (\bm{k})\sigma_\alpha$, 
where $I$, $\sigma_\alpha$, and $h_\alpha$ 
($\alpha=x,y,z$) are respectively the $2\times 2$ identity matrix, 
Pauli matrices, and functions of $\bm{k}$ characterizing materials. 
The two bands cross each other at a WN of $\bm{k}=\bm{K}$ when 
$h_\alpha (\bm{K})=0$. This can happen in three dimensions (3D)
because three conditions match with three variables, and the 
level repulsion principle can at most shift the WNs. 
Moreover, WNs must come in pairs with opposite chirality according 
to the no-go theorem \cite{P}, and the band inversion occurs 
between two paired WNs, resulting in the topologically protected 
surface states and accompanying Fermi arcs on crystal surfaces. 
The only way to destroy a WSM is the merging of two WNs 
of opposite chirality or via superconductivity \cite{W10}.  

How does the above picture based on the lattice translational symmetry 
change when disorders are presented and the lattice momentum is not 
a good quantum number anymore? This is an important question that 
has been investigated intensively with conflicting results 
\cite{Q1,Q2,Q3,Q4,Q5,Q6,Q7,Q8,Q9,Q10,Q11,Q12,Q13,WZD,Fradkin}.
Disorder can greatly modify electronic structures, resulting in the 
well-known Anderson localization. One expects that disorder has 
much more interesting effects to a WSM than that to a normal metal. 
For example, electrons with linear dispersion relations around the WNs 
(Dirac nodes) are governed by the effective Weyl (massless Dirac) equation. 
Weyl electrons cannot be confined by any potential due to the Klein paradox 
\cite{KP}. Early theoretical studies ignored internode scattering and 
predicted that the WSM phase featured by vanishing density of states (DOS) 
at WNs is robust against weak disorder and undergoes a direct quantum phase 
transition to the diffusive metal (DM) phase as disorder increases 
\cite{Fradkin,Q1,Q6,Q7}. The divergence of the bulk state localization 
length at the WSM-DM transition was conjectured \cite{Q1,Q2} and was used 
in recent numerical studies \cite{Q3,Q10,Q11} to support disordered WSMs 
in a wide range of disorder and direct WSM-DM transitions \cite{note1}. 
However, a real WSM has at least two WNs of opposite chirality, and 
disorder can mix two nodes by internode scattering so that the Anderson 
localization can happen as shown in the disordered graphene \cite{yanyang}. 
Therefore, the applicability of the direct WSM-DM transition conjectured by 
theories of a single WN \cite{Fradkin,Q1,Q6,Q7} for real disordered WSMs 
is questionable. The predicted vanishing DOS at WNs have also attracted many 
numerical studies \cite{Q2,Q3,Q5,Q9}, and recent works concluded that 
zero DOS cannot exist at nonzero disorder due to rare region effects 
and no WSM phase is allowed at an arbitrary weak disorder if zero DOS 
at WNs is demanded \cite{Q4,Q5}. 

Strictly speaking, because the lattice momentum is not a good quantum 
number in a disordered WSM, $\bm{k}$-space is only an approximate 
language although the concepts of band and DOS are still accurate. 
Thus, the validity of DOS $\rho(E)\propto E^2$ from 3D linear 
dispersion relations as a signature of disordered WSMs is doubtful. 
The distinct property of a WSM is the existence of topologically 
protected surface states that do not necessarily rely on the 
linear crossing of two bands and zero DOS at WNs, and should be 
robust against disorder, at least against the weak one. Therefore, 
a disordered WSM is defined as a bulk metal with topologically 
protected surface states in this work. Since both the WSM and DM 
are bulk metals, bulk states of them are extended and no 
theoretical basis supports the hypothesis of the divergence of 
localization length at the WSM-DM transition. Focusing on the 
previously proposed quantum critical point between the WSM and DM 
phases \cite{Q3,Q10,Q11}, we show that the so-called direct WSM-DM 
transition actually corresponds to two quantum phase transitions and
a narrow Chern insulator (CI) phase exists between the two 
distinct metallic phases. The critical exponent of localization 
length takes the value of 3D Gaussian unitary ensemble of the 
conventional Anderson localization transition \cite{C1,C2,C3,C4}. 
Nontrivial topological nature of the CI phase is confirmed 
by Hall conductance calculations that show well-defined quantized 
plateaus in the bulk insulating phase. Furthermore, the disorder-induced 
various plateau-plateau transitions between different quantized 
values of Hall conductance can be well explained by the 
self-consistent Born approximation (SCBA). 

In order to compare directly with previous studies, we consider a 
tight-binding Hamiltonian on a cubic lattice of unity lattice 
constant that was used in Refs. \cite{W2,Q10}, 
\begin{equation}
\begin{split}
&H_0=\sum_i m_zc_i^\dagger\sigma_z c_i - \sum_i \bigg[\frac{m_0}{2}(c_{i+
\hat{x}}^\dagger \sigma_z c_{i}+ c_{i+\hat{y}}^\dagger \sigma_z c_{i})\\ 
&+\frac{t}{2}(c_{i+\hat{z}}^\dagger\sigma_z c_{i}+ ic_{i+\hat{x}}^\dagger 
\sigma_x c_i + ic_{i+\hat{y}}^\dagger \sigma_y c_{i})  + \text{H.c.}\bigg], 
\end{split}
\label{e1}
\end{equation}
where $c_{i}^\dagger=(c_{i,\uparrow}^\dagger,c_{i,\downarrow}^\dagger)$ 
and $c_{i}$ are electron creation and annihilation operators at site $i$. 
$\hat{x}$, $\hat{y}$, $\hat{z}$ are unit lattice vectors in $x$, $y$, $z$ 
direction, respectively. $\sigma_{x,y,z}$ are Pauli matrices for spin. 
The Hamiltonian Eq.~(\ref{e1}) can be block diagonalized in the 
momentum space as $H_0=\sum_{\bm{k}}c_{\bm{k}}^\dagger \mathcal{H}_0(\bm{k}) 
c_{\bm{k}}$, where $\mathcal{H}_0(\bm{k})=(m_z-t\cos k_z)\sigma_z
-m_0(\cos k_x+\cos k_y)\sigma_z+t(\sin k_x\sigma_x+ \sin k_y \sigma_y)$. 
The dispersion relation of the Hamiltonian is $\varepsilon_\pm (\bm{k}) 
=\pm \sqrt{\Delta(\bm{k})^2 + t^2(\sin^2 k_x + \sin^2 k_y)}$ with 
$\Delta(\bm{k}) = m_z - t \cos k_z - m_0(\cos k_x + \cos k_y)$. 
In this study, $m_0=2.1t$, identical to that in Ref. \cite{Q10}, is used. 
$m_z$ is the tunable variable to control different phases \cite{mz}. 
The WSM phase requires $\Delta(\bm{k})=0$ at $k_{x,y}=0$ or $\pm \pi$, 
and the model supports various phases \cite{Q10,phase1} at zero Fermi 
energy $E_F=0$. In order to study the disorder effect, a spin-resolved 
on-site disorder is included in the model, 
\begin{equation}
H=H_0+\sum_{i,
\sigma} c_{i,\sigma}^\dagger V_{i,\sigma} c_{i,
\sigma},
\label{e2}
\end{equation}
where $\sigma=\uparrow$ or $\downarrow$ and $\{V_{i,\sigma}\}$ 
are uniformly distributed within $[-W/2,W/2]$. 
Here both $H$ and $H_0$ do not have time-reversal symmetry, and 
$\overline{V_{i,\sigma}}=0$ and $\overline{V_{i,\sigma}V_{i',\sigma'}}
=W^2\delta_{i,i'}\delta_{\sigma,\sigma'}/12$ with the bar denoting 
ensemble average over different configurations. 
According to the Fermi golden rule, the internode and intranode 
scattering around the WNs have the same rate of 
\begin{equation}
\Gamma_{\rm inter}=\Gamma_{\rm intra}=\frac{\pi W^2\rho(E_F)}{24\hbar},
\label{sr}
\end{equation}
where $\rho(E_F)$ is the DOS at Fermi energy \cite{sm} and $\rho(0)\neq0$ 
for nonzero disorder. Therefore the two kinds of scatterings are 
equally important in the disordered WSM. Moreover, because $\rho(E_F)$ 
is an increasing function of $|E_F|$ around WNs, the scattering rates 
increases as the Fermi energy shifts away from the WNs. 
\begin{figure}
  \begin{center}
  \includegraphics[width=8.5 cm]{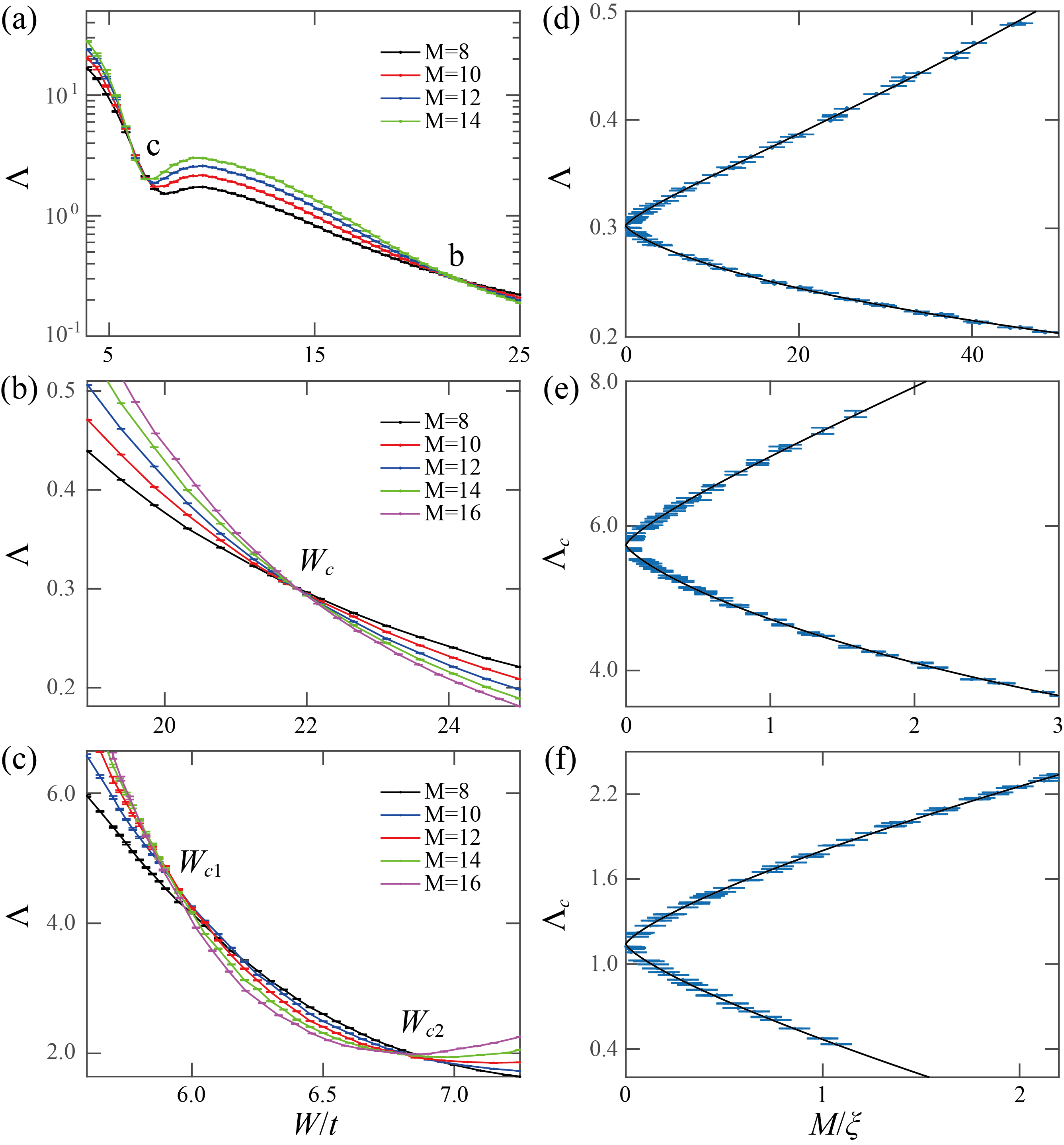}
  \end{center}
  \vspace{-0.2in}
\caption{(color online). \footnotesize{(a) The normalized localization 
length as a function of $W/t$ for various system sizes and with the 
parameters specified in the text. b and c indicate the possible 
quantum phase transition points. (b) and (c) The close-up shots 
of the possible transition regions around b and c in (a). 
(d) The scaling function obtained by collapsing data points around 
the critical point $W_{\rm c}$ in (b) into the smooth curves. 
(e) and (f) The scaling functions obtained from the corrections 
to the single-parameter scaling ansatz by collapsing data points 
around the critical points $W_{c1}$ and $W_{c2}$ in (c) into the 
smooth curves, respectively. }}
  \label{fig1}
  \vspace{-0.2in}
\end{figure}

To investigate various quantum phase transitions in the model, we evaluate 
the localization length by standard transfer matrix method \cite{TM1,TM2}. 
Here we consider a bar of size $M_x\times M_y\times M_z$ with 
$M_z=10^5$ and $M_x=M_y=M$. Periodic boundary conditions are applied 
in both $x$ and $y$ directions in order to eliminate surface effects. 
We fix $m_z=2.19m_0$ in the WSM phase \cite{phase1} since it was reported that 
the system undergoes a WSM-DM transition as disorder increases \cite{Q10}. 
For $E_F=0$, the normalized localization length $\Lambda=\lambda(M)/M$ versus 
$W$ for various $M$ is shown in Fig.~\ref{fig1}(a). Very similar to early 
studies \cite{Q3,Q10,Q11}, two phase transition points b and c of  
$d\Lambda/dM=0$ seem appear. 
Zooming in on these transition regions, the normalized localization length 
are shown in Figs. \ref{fig1}(b) and \ref{fig1}(c) for b and c, respectively. 
Apparently, the normalized localization length curves of different $M$ 
cross at a single critical disorder $W_c$ in Fig. \ref{fig1}(b) that 
separates a region of $d\Lambda/dM>0$ of a metallic phase for $W<W_c$ 
from a region of $d\Lambda/dM<0$ of an insulating phase for $W>W_c$. 
However, there is a narrow insulating phase characterized by $d\Lambda/dM<0$ 
for $W_{c1}<W<W_{c2}$ around c, separating two distinct metallic phases ($d
\Lambda/dM>0$ for $W<W_{c1}$ and $W>W_{c2}$), as shown in Fig. \ref{fig1}(c).

To substantiate the criticality of transitions occurring at $W=W_c, W_{c1},
W_{c2}$, we employ the finite size scaling analysis for these bulk state localization lengths. 
For the transition at b, the single-parameter scaling hypothesis is applied 
as $\Lambda=f(M/\xi)$, where $\xi\sim |W-W_{c}|^{-\nu}$ diverges at the 
transition point. The scaling functions from both metallic (upper branch) 
and insulating (lower branch) sides are shown in Fig. \ref{fig1}(d). 
The perfect collapse of the data points in Figs.~\ref{fig1}(b) into 
the smooth curves supports our claim of the quantum phase transition. 
The analysis yields $W_{c}/t= 21.81\pm 0.02$ and $\nu=1.31\pm 0.02$, 
consistent with the previous numerical and experimental results \cite{C1,
C2,C3,C4} for 3D Gaussian unitary ensemble. For the quantum phase transitions 
at critical points $W_{c1}$ and $W_{c2}$ shown in Fig.~\ref{fig1}(c), 
the crossing of different curves is less perfect as it often happens in 
3D systems when the system size is limited by the computer resources.
We therefore follow the more accurate analysis used in Ref. \cite{CS} 
to include the contributions of the most important irrelevant 
parameter to the scaling function 
\begin{equation}
\Lambda=F(\psi M^{1/\nu},\phi M^\mu),
\label{e4}
\end{equation}
where $\psi$ is the relevant scaling variable with $\nu>0$ 
and $\phi$ is the irrelevant scaling variable with $\mu<0$.  
Using $\nu=1.30$ for the 3D Gaussian unitary class and by 
minimizing $\chi^2$, we fit the data points around the two 
transition points shown in Fig.~\ref{fig1}(c) to the scaling 
function Eq.~(\ref{e4}) \cite{sm}. 
Indeed, the perfect scaling curves in Figs.~\ref{fig1}(e) and \ref{fig1}(f) 
with $W_{c1}/t=5.81\pm 0.06$ and $W_{c2}/t=6.58\pm0.19$ support our analysis. 
The chi square of the two fittings are $\chi^2=78.80$ and $82.49$ with the 
degrees of freedom $N_d=$ 86 and 88 (the number of data points minus the number 
of fitting parameters), respectively. The reduced chi square of the two cases 
are $\chi_{\rm red}^2=\chi^2/N_d=0.92$ and 0.94, quite satisfactory numbers. 
We also calculate the localization length for various $m_z$ 
and $E_F$ in the WSM phase \cite{sm}. 
It is shown that the insulating phase between the two distinct metallic 
phases is generic. A phase diagram is constructed in the $m_z/m_0$-$W/t$ 
plane for $E_F=0$ and will be discussed below. As $E_F$ increases from 
zero energy, the intermediate insulating phase expands initially since 
the internode scattering rate increases with $E_F$ as shown in Eq.~(\ref{sr}). 
Further increase of $E_F$, the linear dispersion relation fails and the 
system becomes a conventional 3D metal with Fermi energy deep inside the conduction band.  

\begin{figure}
  \begin{center}
  \includegraphics[width=8.5 cm]{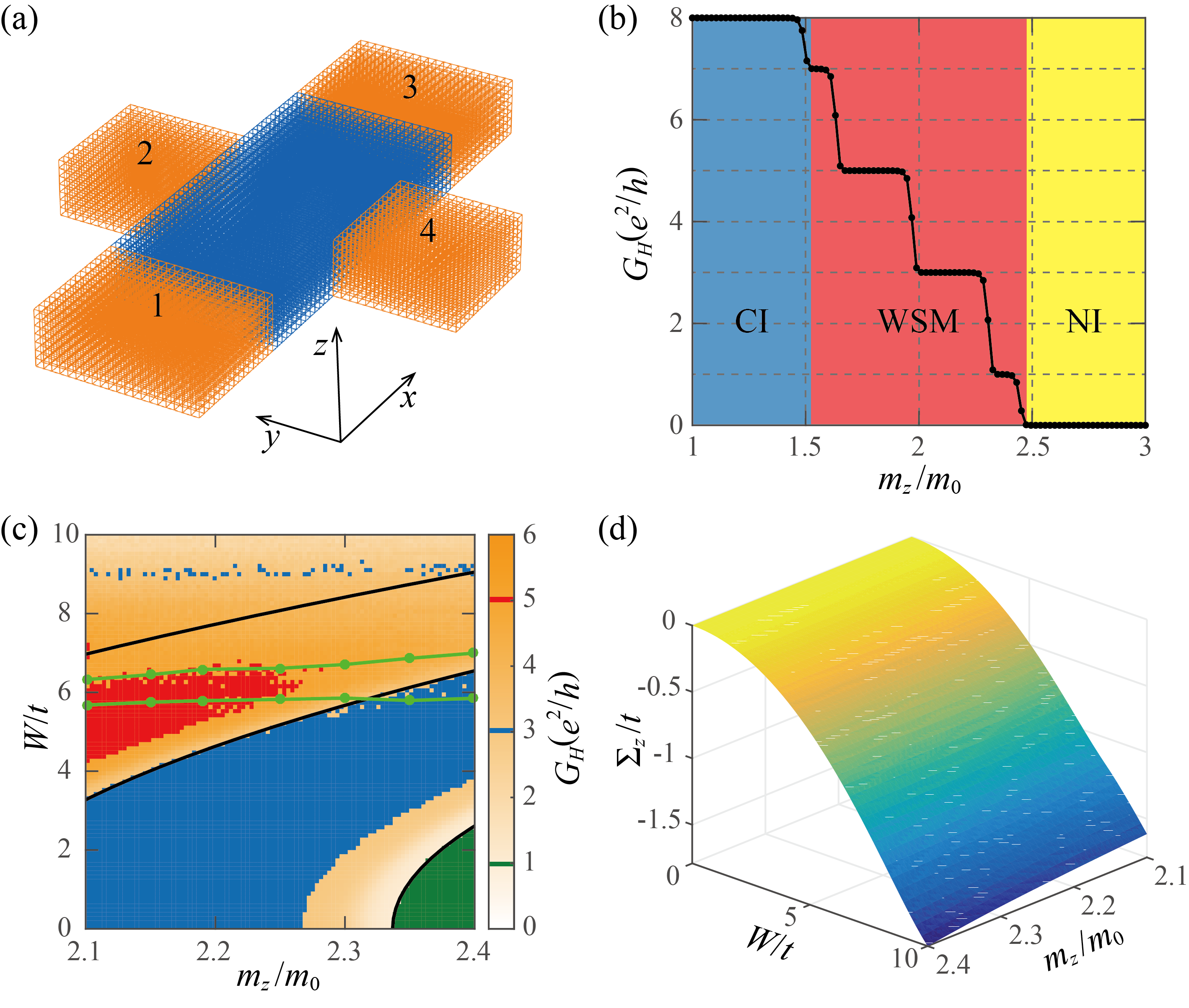}
  \end{center}
  \vspace{-0.2in}
\caption{(color online). \footnotesize{(a) The sketch 
of a four-terminal Hall bar. The blue region is 
described by the Hamiltonian Eq.~(\ref{e2}). 
The four semi-infinite metallic leads are represented 
by the orange parts. (b) The Hall conductance as a 
function of $m_z/m_0$ for the clean system. 
The shallow blue, red, and yellow regions mark the CI, 
WSM, and normal insulator (NI) phases, respectively. 
(c) The density plot of Hall conductance in the 
$m_z/m_0$-$W/t$ plane for the disordered system. 
The three black lines (from bottom to top) are 
plateau-plateau transition lines obtained 
from the SCBA for $n=1$, 2, 3 in Eq.~(\ref{e7}). 
The two green lines enclose the CI phase region 
according to the localization length calculations. 
(d) The $\Sigma_z$ component of the self-energy obtained 
from the SCBA as a function of $m_z/m_0$ and $W/t$.}} 
  \label{fig2}
  \vspace{-0.2in}
\end{figure}

In order to investigate the chiral surface states and topological 
nature of the intermediate insulating phase identified above, we calculate 
the quantum conductance of a four-terminal Hall bar of size 
$80\times40\times8$ marked by blue color in Fig.~\ref{fig2}(a). 
The bar is described by the Hamiltonian Eq.~(\ref{e2}), and the periodic 
boundary condition is applied in the $z$ direction while the 
open boundary condition is applied in the $x$ and the $y$ directions. 
Four semi-infinite metallic leads marked by orange color are 
connected to the bar as shown in Fig.~\ref{fig2}(a). 
One can view the system as coupled multiple two-dimensional 
subsystems of $\mathcal{H}_0(\bm{k})=\sum_{k_z}h_{k_z}(k_x,k_y)$ with 
$k_z=2\pi n/8$, where the integer $n\in[-4,4)$ labels allowed $k_z$ 
within the first Brillouin zone (BZ). For $k_z\neq K_z$ (WNs 
\cite{phase1}), two-dimensional Hamiltonians $h_{k_z}(k_x,k_y)$ are 
gapped whose Chern number $C(k_z)$ is $C(|k_z|<| K_z|)=1$ and 
$C(|k_z|>|K_z|)=0$ for $2m_0-t<m_z<2m_0+t$ \cite{Q10,phase1}. 
Thus, a chiral surface state must exist for each allowed $k_z\in(-
|K_z|,|K_z|)$, and contribute a quantized Hall conductance of $e^2/h$. 
Therefore, the total Hall conductance from the surface states is 
$G_H=e^2/h\sum_{k_z}C(k_z)=e^2|K_z|M_z/(h\pi)$. 
The Hall conductivity is $\sigma_H=G_HM_x/(M_yM_z)=e^2|K_z|M_x/(h\pi M_y)$. 
Moreover, in the CI phase, $C(k_z)=\pm 1$ for all the $k_z$ \cite{Q10,phase1}. 
Thus, the Hall conductance is $G_H=\pm M_ze^2/h$. 

The Hall conductance in the absence of contact resistance can be 
calculated from the formula \cite{HC}  
\begin{equation}
G_H\equiv I_{13}/V_{24}=(e^2/h)(T_{12}-T_{14}),
\end{equation}
where $T_{ij}$ is the transmission coefficient from lead $j$ to 
lead $i$, and current $I_i$ in lead $i$ is given by the 
Landauer-B{\" u}ttiker formalism $I_i=(e^2/h)\sum_{j\neq i}(T_{ji}
V_i-T_{ij}V_j)$ where the voltage on lead $i$ is $V_i$ \cite{L,B}. 
For the clean system, the Hall conductance as a function of $m_z/m_0$
is shown in Fig.~\ref{fig2}(b). As expected, the Hall conductances in  
the normal insulator (NI) and CI phases \cite{phase1} are respectively 
$0$ and $8e^2/h$. In the WSM phase, there are various plateau-plateau 
transitions between quantized Hall conductances $G_H\in(0,8e^2/h)$. 
Because the change of $m_z$ shifts WN positions of  
$\bm{K}=(0,0,\pm\cos^{-1}(m_z/t-2m_0/t))$ \cite{phase1}, the transition 
from $(2n+1)e^2/h$-plateau to $(2n-1)e^2/h$-plateau occurs whenever 
$m_z=2m_0+t\cos(2\pi n/8)$, where $n=1$, 2, 3 in the current case.  

The density plot of Hall conductance (ensemble average over 20 configurations) 
at $E_F=0$ in the $m_z/m_0$-$W/t$ plane is shown in Fig.~\ref{fig2}(c). 
For $m_z/m_0\in[2.1,2.4]$, the clean system is a WSM whose Hall 
conduction at WNs is from the surface states and is quantized at 
a value determined by $m_z$ as mentioned early. 
Interestingly, at a fixed $m_z$ (along a vertical line in Fig. 
~\ref{fig2}(c)), the Hall conductance can jump from one quantized 
value into another as disorder increases.  
In order to understand these transitions, we use the SCBA to see how 
the disorder modifies the model parameters \cite{Q10,SCBA1,SCBA2}. 
The self-energy at the Fermi energy due to the disorder is 
\begin{equation}
\Sigma(m_z,W)=\frac{W^2}{12S_{\rm BZ}}\int_{\rm BZ}
d^3 \bm{k}[E_F+i0^{+}-\mathcal{H}
(\bm{k},m_z,W)]^{-1},
\label{e6}
\end{equation}
where $S_{\rm BZ}=8\pi^3$ is the volume of the first BZ and 
$\mathcal{H}(\bm{k},m_z,W)=\mathcal{H}_0(\bm{k})+\Sigma(m_z,W)$ is the 
effective Hamiltonian. For $E_F=0$, one has $\Sigma=\Sigma_z\sigma_z$ 
since $\mathcal{H}$ has the particle-hole symmetry \cite{S}. 
The dispersion relation of the effective Hamiltonian $\mathcal{H}$ 
is then $\widetilde{\varepsilon}_\pm (\bm{k})=\pm 
\sqrt{[\Delta(\bm{k})+\Sigma_z]^2 + t^2(\sin^2 k_x+\sin^2 k_y)}$. 
Eq.~(\ref{e6}) is solved numerically and $\Sigma_z(m_z,W)$ is 
shown in Fig.~\ref{fig2}(d). Apparently, $\Sigma_z(m_z,W)<0$ and 
is a monotonically decreasing function of $W$. Consequently, the 
modified mass term $\widetilde{m}_z=m_z+\Sigma_z$ decreases and 
the WNs at $\bm{K}=(0,0,\pm\cos^{-1}(\widetilde{m}_z/t-2m_0/t))$ 
are shifted towards the BZ boundary as $W$ increases. 
The plateau-plateau transitions occur at
\begin{equation}
\widetilde{m}_z(m_z,W)=2m_0+t\cos(2\pi n/M_z),
\label{e7}
\end{equation} 
which are plotted as three black curves in Fig~\ref{fig2}(c) for $n=1$, 
2, 3 (from bottom to top), respectively. They separate different plateaus. 
The system becomes a DM at strong disorder (about $W/t>7$), where the 
SCBA is not expected to work and no quantized Hall conductance is observed. 

Our results from localization length and quantum transport calculations are 
summarized in the phase diagram and the density plot of Hall conductance
in the $m_z/m_0$-$W/t$ plane for $E_F=0$ as shown in Fig.~\ref{fig2}(c). 
Only those $m_z$, at which the clean system is in the WSM phase and was 
reported to undergo the WSM-DM transition as disorder increases \cite{Q10}, 
are considered. The two green curves are the boundaries of 
the DM/CI phases (upper line) and CI/WSM phases (lower line). 
The narrow CI phase region separates the WSM phase from the DM phase. 
The CI phase is inferred from the fact that all bulk states are localized 
according to the localization length calculations while the Hall conductance 
of a finite bar is nonzero and takes several quantized values (red for 5, blue 
for 3, and green for 1 in units of $e^2/h$), as shown in Fig. \ref{fig2}(c). 
The WSM phase is defined as bulk metallic states (extended wavefunctions) 
with edge conducting channels while the DM phase has bulk metallic states 
without edge conducting channels. Both the CI and WSM phases can have well 
quantized Hall conductance (red, blue, and green regions in Fig. 
\ref{fig2}(c)) while quantized Hall conductance is absent in the DM phase. 

The generality of the no direct WSM-DM transition can be understood 
from the following reasoning. In order to have a direct WSM-DM 
transition, WNs and topologically protected surface states should be 
destroyed simultaneously. However, the two events are not exactly 
the same although they are related. The topologically protected surface 
states are due to nonzero band Chern numbers of two-dimensional  
slices between the two WNs. 
In general, disorder pushes the two WNs away from each other and towards 
the BZ boundary (as elaborated by the SCBA) where they can merge.
As a result, the WNs are destroyed while the nonzero band Chern numbers 
of two-dimensional slices survive, resulting in the intermediate CI phase.  
Whether disorder can pull two paired WNs together and towards the BZ center 
so that the WNs and band Chern numbers can simultaneously be destroyed 
is an open question.

In conclusion, we show that the claimed direct transition from a 
WSM to a DM do not exist under uncorrelated on-site disorder due 
to non-negligible internode scattering. Instead, there exists a 
intermediate CI phase that separates a WSM phase from a DM phase. 
Namely, there are actually two quantum phase transitions between the 
disordered WSM and the DM: One is from the WSM to the CI, and the 
other is from the CI to the DM. 
The critical exponent of $\nu\simeq 1.3$ suggests that the two 
transitions belong to the same universality class of the 3D Gaussian 
unitary ensemble of the conventional Anderson localization transition. 
The intermediate CI phase persists and expands at weak disorder as 
the Fermi energy slightly shifts away from the WNs. Our results do  
not dependents on specific choices of lattice model since the 
analysis based on low-energy effective Weyl Hamiltonians is general.  

\begin{acknowledgements}
{\it Acknowledgements.---}We would like to thank Chuizhen Chen and 
Ryuichi Shindou for helpful discussion. This work was supported by 
National Natural Science Foundation of China Grant No. 11374249 
and Hong Kong Research Grants Council Grants  No. 163011151 and 16301816. 
XSW acknowledges support from UESTC. 
\end{acknowledgements}

\widetext
\pagebreak

\begin{center}
\textbf{Supplemental Material for A Generic Phase between Disordered Weyl Semimetal and Diffusive Metal}
\end{center}
\setcounter{equation}{0}
\setcounter{figure}{0}
\setcounter{table}{0}
\setcounter{page}{1}
\makeatletter
\renewcommand{\theequation}{S\arabic{equation}}
\renewcommand{\thefigure}{S\arabic{figure}}
\renewcommand{\bibnumfmt}[1]{[S#1]}
\renewcommand{\citenumfont}[1]{S#1}

\subsection{Internode and intranode scattering rates}\label{scattering}

\begin{figure}[b]
  \begin{center}
  \includegraphics[width=9 cm]{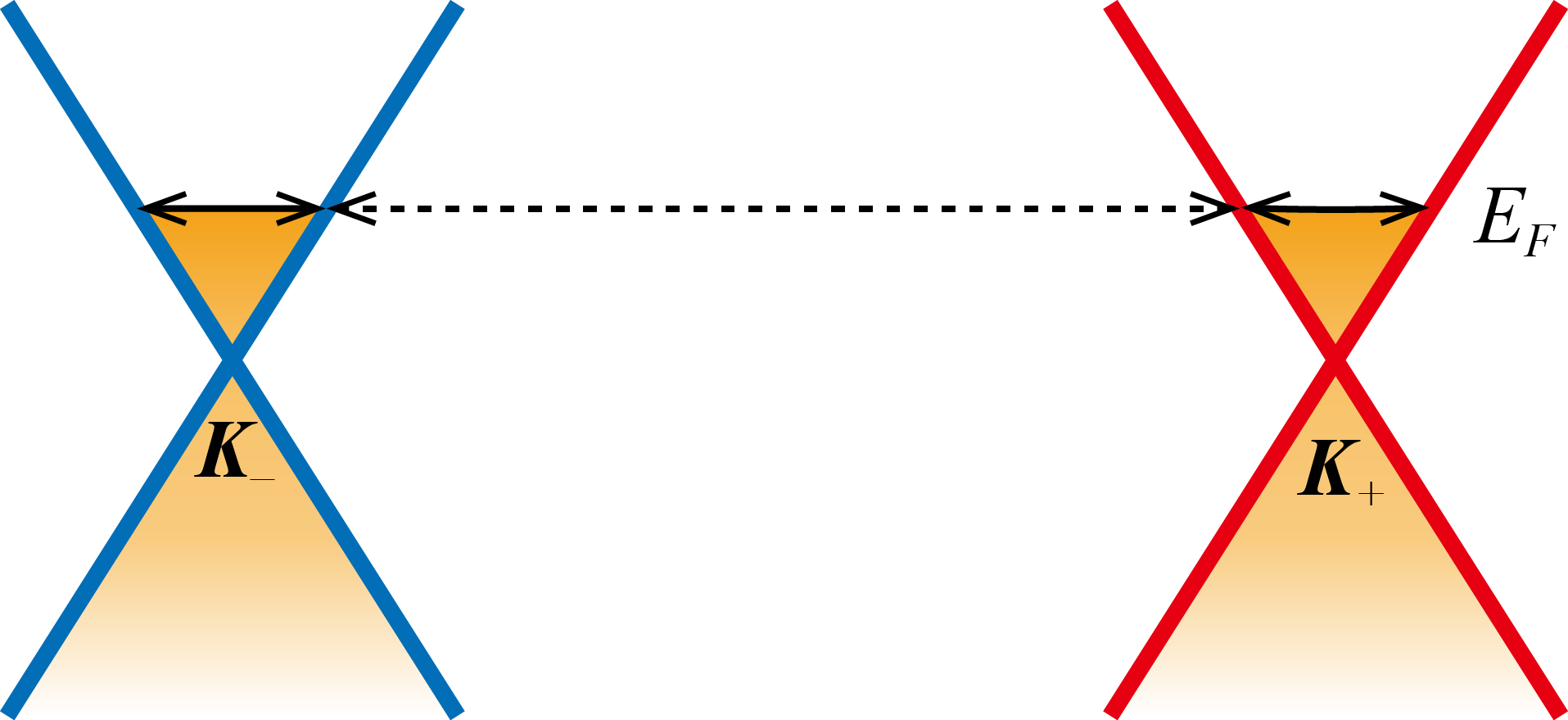}
  \end{center}
  \vspace{-0.2in}
\caption{\footnotesize{Schematic diagram of internode scattering 
represented by the dashed arrow and intranode scattering represented 
by the solid arrows.}}
  \label{s1}
  \vspace{-0.2in}
\end{figure}

The rates of internode and intranode scatterings caused by uncorrelated 
on-site disorder are derived from low-energy effective Weyl 
Hamiltonians in this section. For the model parameters $m_z\in (2m_0-t,2m_0+t)$ 
studied in the manuscript, the clean system supports a pair of Weyl nodes (WNs) at
\begin{equation}
\bm{K}_\pm = \left(0,0,\pm \cos^{-1} \frac{m_z-2m_0}{t}\right).
\end{equation}
The low-energy effective Weyl Hamiltonians (to the first order in 
the momentum deviation $\bm{q}=\bm{k}-\bm{\bm{K}}_\pm$) around 
the WNs $\bm{K}_\pm$ can be obtained from the Taylor expansion as
\begin{equation}
\mathcal{H}_\pm(\bm{q})= \sum_{\alpha=x,y,z}\hbar v^\pm_\alpha 
q_\alpha\sigma_\alpha,
\end{equation}
where the Fermi velocities are $v_x^\pm=v_y^\pm=t/\hbar$ and 
$v_z^\pm=\pm\sqrt{t^2-(m_z-2m_0)^2}/\hbar$. The energy bands 
of the Weyl Hamiltonians are $\pm E_{\bm{q}}=\pm\sqrt{
\sum_\alpha \hbar^2v_\alpha^{\pm2}q_\alpha^2}$ whose  conduction ($c$) 
and valence ($v$) band eigenstates are 
\begin{equation}
|c,\bm{K}_\pm+\bm{q}\rangle=
\left(
\begin{matrix}
\frac{\hbar v_z^\pm q_z + \sqrt{\sum_\alpha \hbar^2 v_\alpha^{\pm 2} 
q_\alpha^2}}{\sqrt{2\sum_\alpha \hbar^2 v_\alpha^{\pm 2} q_\alpha^2 
+ 2\hbar v_z^\pm q_z\sqrt{\sum_\alpha \hbar^2v_\alpha^{\pm 2} q_\alpha^2}}} \\
\frac{\hbar v_x^\pm q_x+i\hbar v_y^\pm q_y}{\sqrt{2\sum_\alpha\hbar^2 
v_\alpha^{\pm 2} q_\alpha^2 + 2\hbar v_z^\pm q_z\sqrt{\sum_\alpha
\hbar^2 v_\alpha^{\pm 2} q_\alpha^2}}}
\end{matrix}\right),
\;\;
|v,\bm{K}_\pm+\bm{q}\rangle=
\left(
\begin{matrix}
\frac{\hbar v_z^\pm q_z - \sqrt{\sum_\alpha \hbar^2 v_\alpha^{\pm 2} 
q_\alpha^2}}{\sqrt{2\sum_\alpha \hbar^2 v_\alpha^{\pm 2} q_\alpha^2 
- 2\hbar v_z^\pm q_z\sqrt{\sum_\alpha \hbar^2v_\alpha^{\pm 2} q_\alpha^2}}}\\
\frac{\hbar v_x^\pm q_x-i\hbar v_y^\pm q_y}{\sqrt{2\sum_\alpha\hbar^2 
v_\alpha^{\pm 2} q_\alpha^2 - 2\hbar v_z^\pm q_z\sqrt{\sum_\alpha\hbar^2 
v_\alpha^{\pm 2} q_\alpha^2}}}
\end{matrix}\right).
\end{equation}
To be concrete and without losing generality, we fix the Fermi energy  
in the conduction band $E_F=E_{\bm{q}}$ as shown in Fig. \ref{s1}. 
In order to shorten the notation, we denote $a_{\bm{q}}^\pm=\cos\frac{\theta
_{\bm{q}}^\pm}{2}$, $b_{\bm{q}}^\pm=\sin\frac{\theta_{\bm{q}}^\pm}{2}$, 
$\cos\theta_{\bm{q}}^\pm = \frac{\hbar v_z^\pm q_z}{E_F}$, and 
$\tan\phi_{\bm{q}}^\pm=\frac{v_y^\pm q_y}{v_x^\pm q_x}$, 
so that the eigenstates with the Fermi energy can be expressed as
\begin{equation}
|c,\bm{K}_\pm+\bm{q}\rangle=
\left(\begin{matrix}
a_{\bm{q}}^\pm\\
b_{\bm{q}}^\pm e^{i\phi_{\bm{q}}^\pm}
\end{matrix}\right).
\end{equation}

In the presence of disorder, the transition rate from an initial state 
$|\bm{k}\rangle$ to a final state $|\bm{k}'\rangle$ caused by elastic 
scattering is given by the Fermi golden rule
\begin{equation}
\Gamma_{\bm{k},\bm{k}'}=\frac{2\pi}{\hbar}\overline{|\langle\bm{k}'|
V|\bm{k}\rangle|^2}\delta(E_{\bm{k}'}-E_{\bm{k}}),
\end{equation}
where $V$ encodes the disorder and the bar denotes ensemble average 
over different configurations. For the uncorrelated on-site disorder 
used in the manuscript
\begin{equation}
V=\sum_{j,
\sigma} c_{j,\sigma}^\dagger V_{j,\sigma} c_{j,
\sigma},
\end{equation}
the total scattering processes consist of two parts: the internode 
scattering and intranode scattering that are schematically shown 
in Fig. \ref{s1}. According to the Fermi golden rule, their scattering 
rates are respectively
\begin{equation}
\Gamma_{\rm inter}=\sum_{\bm{q}'}\frac{2\pi}{\hbar}\overline{|\langle c,
\bm{K}_{\mp}+\bm{q}'|V|c,\bm{K}_\pm+\bm{q}\rangle|^2}\delta(E_{\bm{q}'}-E_F),\\
\label{inter}
\end{equation}
\begin{equation}
\Gamma_{\rm intra}=\sum_{\bm{q}'}\frac{2\pi}{\hbar}\overline{|\langle c,
\bm{K}_{\pm}+\bm{q}'|V|c,\bm{K}_\pm+\bm{q}\rangle|^2}\delta(E_{\bm{q}'}-E_F).
\label{intra}
\end{equation}
Here the internode scattering amplitudes are 
\begin{equation}
\begin{split}
\langle c,\bm{K}_\mp+\bm{q}'|V|c,\bm{K}_\pm+\bm{q}\rangle
&=\frac{1}{N}\sum_j\left(a_{\bm{q}'}^\mp,b_{\bm{q}'}^\mp e^{-i\phi_{\bm{q}'}^\mp}
\right)\left(\begin{matrix}
V_{j,\uparrow} & \\
& V_{j,\downarrow} 
\end{matrix}\right)
\left(\begin{matrix}
a_{\bm{q}}^\pm\\
b_{\bm{q}}^\pm e^{i\phi_{\bm{q}}^\pm}
\end{matrix}\right)e^{i(\bm{K}_\pm+\bm{q}-\bm{K}_\mp-\bm{q}')\cdot\bm{r}_j}\\
&=\frac{1}{N}\sum_j\left(a_{\bm{q}'}^\mp a_{\bm{q}}^\pm V_{j,\uparrow}+
b_{\bm{q}'}^\mp b_{\bm{q}}^\pm e^{i(\phi_{\bm{q}}^\pm-\phi_{\bm{q}'}^\mp)} 
V_{j,\downarrow} \right)e^{i(\bm{K}_\pm+\bm{q}-\bm{K}_\mp-\bm{q}')\cdot\bm{r}_j}\\
&=\left(a_{\bm{q}'}^\mp a_{\bm{q}}^\pm V_{\bm{K}_\pm+\bm{q}-\bm{K}_\mp-\bm{q}',
\uparrow}+b_{\bm{q}'}^\mp b_{\bm{q}}^\pm e^{i(\phi_{\bm{q}}^\pm-\phi_{\bm{q}'}^\mp)} 
V_{\bm{K}_\pm+\bm{q}-\bm{K}_\mp-\bm{q}',\downarrow} \right),
\end{split}
\end{equation}
and the intranode scattering amplitudes are
\begin{equation}
\begin{split}
\langle c,\bm{K}_\pm+\bm{q}'|V|c,\bm{K}_\pm+\bm{q}\rangle
&=\frac{1}{N}\sum_j\left(a_{\bm{q}'}^\pm,b_{\bm{q}'}^\pm e^{-i\phi_{\bm{q}'}^\pm}
\right)\left(\begin{matrix}
V_{j,\uparrow} & \\
& V_{j,\downarrow} 
\end{matrix}\right)
\left(\begin{matrix}
a_{\bm{q}}^\pm\\
b_{\bm{q}}^\pm e^{i\phi_{\bm{q}}^\pm}
\end{matrix}\right)e^{i(\bm{q}-\bm{q}')\cdot\bm{r}_j}\\
&=\frac{1}{N}\sum_j\left(a^{\pm}_{\bm{q}'}a^{\pm}_{\bm{q}} 
V_{j,\uparrow}+b^{\pm}_{\bm{q}'}b^{\pm}_{\bm{q}} e^{i(\phi_{\bm{q}}^\pm-
\phi_{\bm{q}'}^\pm)} V_{j,\downarrow} \right)e^{i(\bm{q}-\bm{q}')\cdot\bm{r}_j}\\
&=\left(a^{\pm}_{\bm{q}'}a^{\pm}_{\bm{q}} V_{\bm{q}-\bm{q}',\uparrow}+
b^{\pm}_{\bm{q}'}b^{\pm}_{\bm{q}} e^{i(\phi_{\bm{q}}^\pm-\phi_{\bm{q}'}^\pm)} 
V_{\bm{q}-\bm{q}',\downarrow} \right),
\end{split}
\end{equation}
where $N$ is the total number of lattice sites and  $V_{\bm{k},\sigma}$ is  
the Fourier transform of $V_{j,\sigma}$ as
\begin{equation}
V_{\bm{k},\sigma}=\frac{1}{N}\sum_j V_{j,\sigma}e^{i\bm{k}\cdot \bm{r}_j}.
\end{equation}
The correlation function of $V_{\bm{k},\sigma}$ is 
\begin{equation}
\overline{V_{\bm{k},\sigma},V_{\bm{k}',\sigma'}}=\frac{1}{N^2}\sum_{j,j'}
\overline{V_{j,\sigma}V_{j',\sigma'}}e^{i\bm{k}\cdot \bm{r}_j-i\bm{k}'\cdot 
\bm{r}_{j'}}=\frac{W^2}{12N}\delta_{\bm{k},\bm{k}'}\delta_{\sigma,\sigma'}.
\end{equation}
Substituting these results back into Eqs. (\ref{inter}) and (\ref{intra}), 
we get the internode scattering rate 
\begin{equation}
\begin{split}
\Gamma_{\rm inter}&=\sum_{\bm{q}'}\frac{2\pi}{\hbar}\overline{\left|
\left(a_{\bm{q}'}^{\mp} a_{\bm{q}}^{\pm} V_{\bm{K}_{\pm}+\bm{q}-
\bm{K}_{\mp}-\bm{q}',\uparrow}+b_{\bm{q}'}^{\mp} b_{\bm{q}}^{\pm} 
e^{i(\phi_{\bm{q}}^{\pm}-\phi_{\bm{q}'}^{\mp})} V_{\bm{K}_{\pm}+\bm{q}-
\bm{K}_{\mp}-\bm{q}',\downarrow} \right)\right|^2}\delta(E_{\bm{q}'}-E_F) \\
&=\sum_{\bm{q}'}\frac{2\pi}{\hbar}\frac{W^2}{12N}\left(a_{\bm{q}'}^{\mp2} 
a_{\bm{q}}^{\pm2}+b_{\bm{q}'}^{\mp2} b_{\bm{q}}^{\pm2}\right)\delta(E_{\bm{q}'}-E_F) \\
&=\sum_{\bm{q}'}\frac{\pi W^2}{12\hbar N}\frac{E_F^2-\hbar^2v_z^{\pm2}
q_z'q_z}{E_F^2}\delta(E_{\bm{q}'}-E_F)=\frac{\pi W^2\rho(E_F)}{24\hbar},
\end{split}
\end{equation}
and the intranode scattering rate
\begin{equation}
\begin{split}
\Gamma_{\rm intra}
&=\sum_{\bm{q}'}\frac{2\pi}{\hbar}\overline{\left|\left(a^{\pm}_{\bm{q}'}
a^{\pm}_{\bm{q}} V_{\bm{q}-\bm{q}',\uparrow}+b^{\pm}_{\bm{q}'}b^{\pm}_{\bm{q}} 
e^{i(\phi_{\bm{q}}^\pm-\phi_{\bm{q}'}^\pm)} V_{\bm{q}-\bm{q}',\downarrow} 
\right)\right|^2}\delta(E_{\bm{q}'}-E_F) \\
&=\sum_{\bm{q}'}\frac{2\pi}{\hbar}\frac{W^2}{12N}\left(a^{\pm2}_{\bm{q}'}
a^{\pm2}_{\bm{q}}+b^{\pm2}_{\bm{q}'}b^{\pm2}_{\bm{q}}\right)\delta(E_{\bm{q}'}-E_F)\\
&=\sum_{\bm{q}'}\frac{\pi W^2}{12\hbar N}\frac{E_F^2+\hbar^2v_z^{\pm2}
q_z'q_z}{E_F^2}\delta(E_{\bm{q}'}-E_F)=\frac{\pi W^2\rho(E_F)}{24\hbar}.
\end{split}
\end{equation}
where $\rho(E_F)$ is the density of states.
Therefore, we conclude that the internode and intranode scattering 
rates are identical in Weyl semimetals subject to uncorrelated on-site 
disorder. Moreover, the scattering rates increases with $|E_F|$ since 
the density of states is an increasing function of $|E_F|$.

\subsection{Correction to the single-parameter scaling hypothesis}

Following the more accurate analysis used in Ref. \cite{sCS} to include 
the contributions from the most important irrelevant parameter, 
the scaling function becomes  
\begin{equation}
\Lambda=F(\psi M^{1/\nu},\phi M^\mu),
\end{equation}
where $\psi$ is the relevant scaling variable with $\nu>0$ and $\phi$ is 
the irrelevant scaling variable with $\mu<0$. Under the Taylor expansion 
around the transition point, the scaling function is $\Lambda=\sum_{n=0}^
{n_I}\phi^n M^{n\mu}F_n(\psi M^{1/\nu})$ and 
$F_n(\psi M^{1/\nu})=\sum_{m=0}^{n_R} \psi^m M^{m/\nu}F_{nm}$, where $\psi
=b(W-W_{c})$ and $\phi=c_0+c_1(W-W_{c})$ up to the first order \cite{sCS}. 
One can remove the contributions from the irrelevant scaling variable to 
$\Lambda$ and define the corrected localization length as 
\begin{equation}
\Lambda_{ c}=\Lambda-\sum_{n=1}^{n_I}\phi^n M^{n\mu}F_n(\psi M^{1/\nu}).
\end{equation} 
Then, the corrected localization length follows the scaling law, 
$\Lambda_c=f(M/\xi)$ and $\xi\sim |W-W_{c}|^{-\nu}$. 
In our analysis, we choose $n_I=n_R=2$ and $F_{01}=F_{10}=1$ \cite{sCS}.

\subsection{Localization length for various $\bm{m_z}$}

\begin{figure}
  \begin{center}
  \includegraphics[width=13 cm]{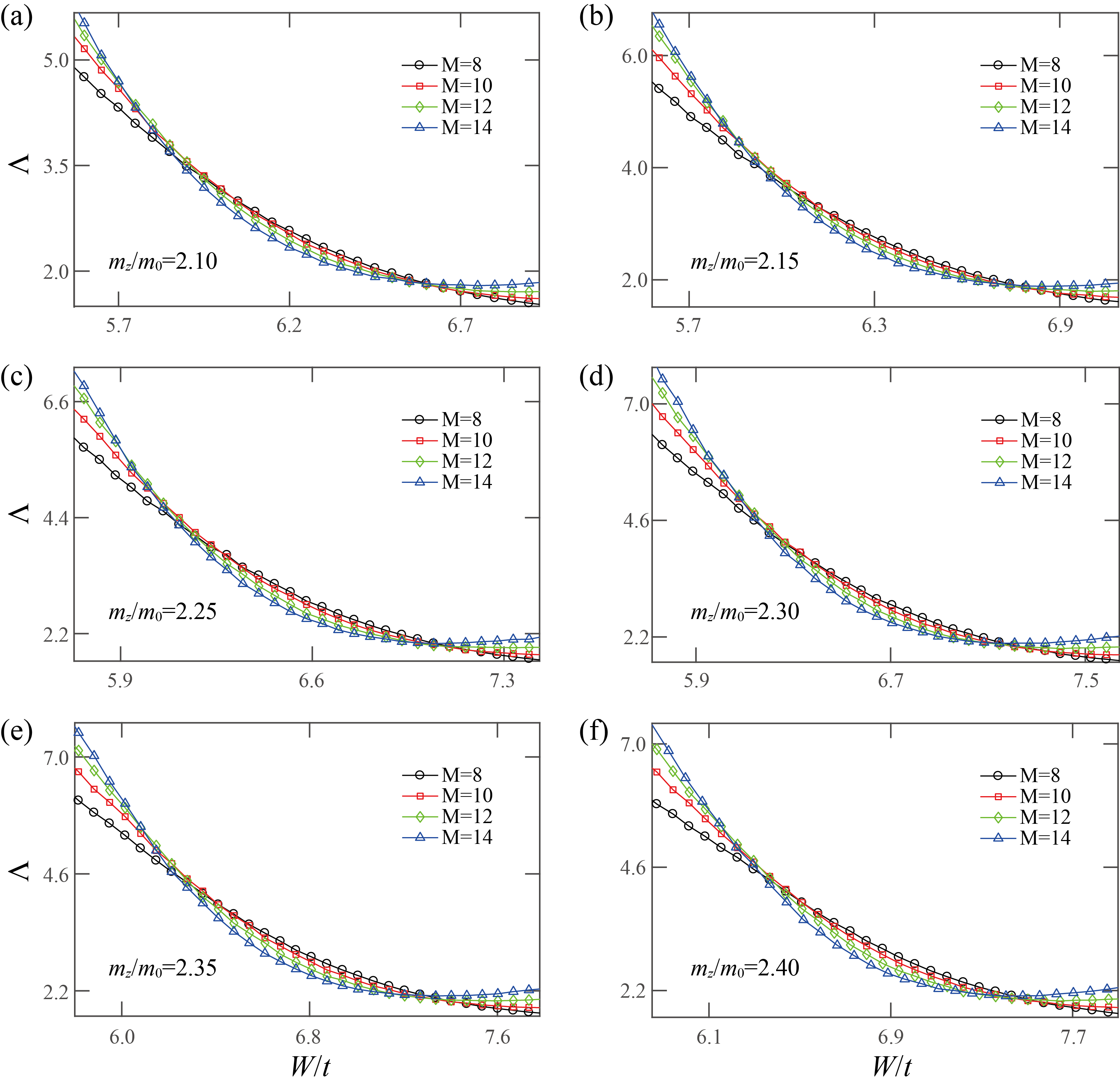}
  \end{center}
  \vspace{-0.2in}
\caption{\footnotesize{(a)-(f)}  Localization length as a function 
of $W/t$ for $m_z/m_0=2.10$, 2.15, 2.25, 2.30, 2.35, 2.40, respectively. 
$E_F=0$ and $m_0=2.1t$ are fixed.}
  \label{s2}
  \vspace{-0.2in}
\end{figure}

In order to show the dependence of the intermediate Chern insulator (CI) 
phase on the model parameter $m_z$ and construct the phase diagram as shown in Fig. 2(c) 
of the manuscript, we calculate the localization length for various $m_z$ 
with $m_0=2.1t$ and $E_F=0$ same as that used in the manuscript. 
The numerical results are shown in Fig. \ref{s2}. Apparently, the 
intermediate CI phase characterized by $d\Lambda/dM<0$ and  quantized nonzero
Hall conductances shown in Fig. 2(c) of the manuscript is generic.

\subsection{Localization length for various $\bm{E_F}$}

\begin{figure}
  \begin{center}
  \includegraphics[width=10 cm]{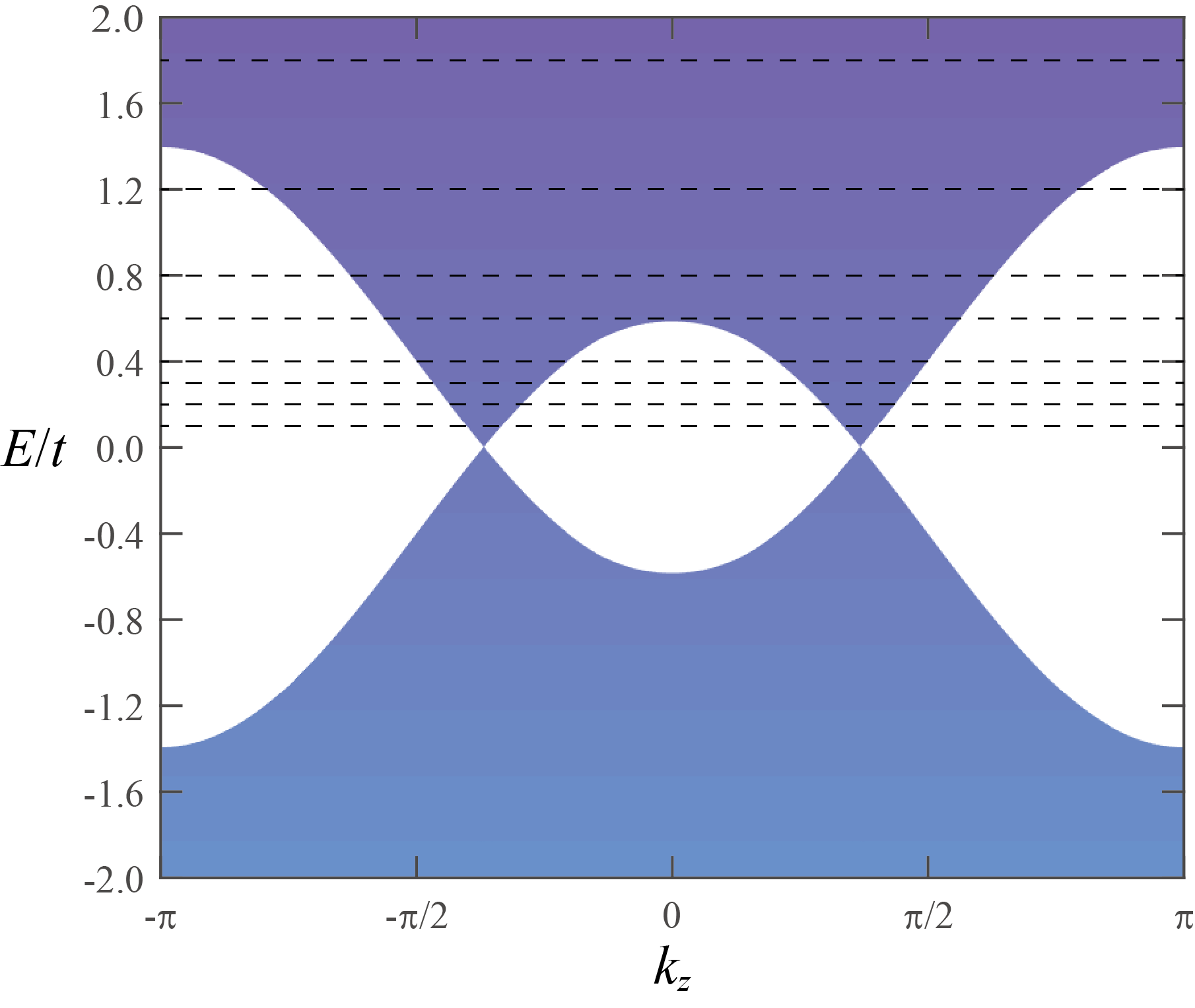}
  \end{center}
  \vspace{-0.2in}
\caption{\footnotesize{Bulk energy bands of the clean system (with the model 
parameters $m_z=2.19m_0$ and $m_0=2.1t$) projected onto the $k_z$-$E$ plane. 
The dashed lines (from down to up) denotes the Fermi energies 
$E_F/t$=0.1, 0.2, 0.3, 0.4, 0.6, 0.8, 1.2, 1.8, respectively, 
that are used for localization length calculations as shown in Fig. \ref{s4}.}}
  \label{s3}
  \vspace{-0.2in}
\end{figure}

\begin{figure}
  \begin{center}
  \includegraphics[width=13 cm]{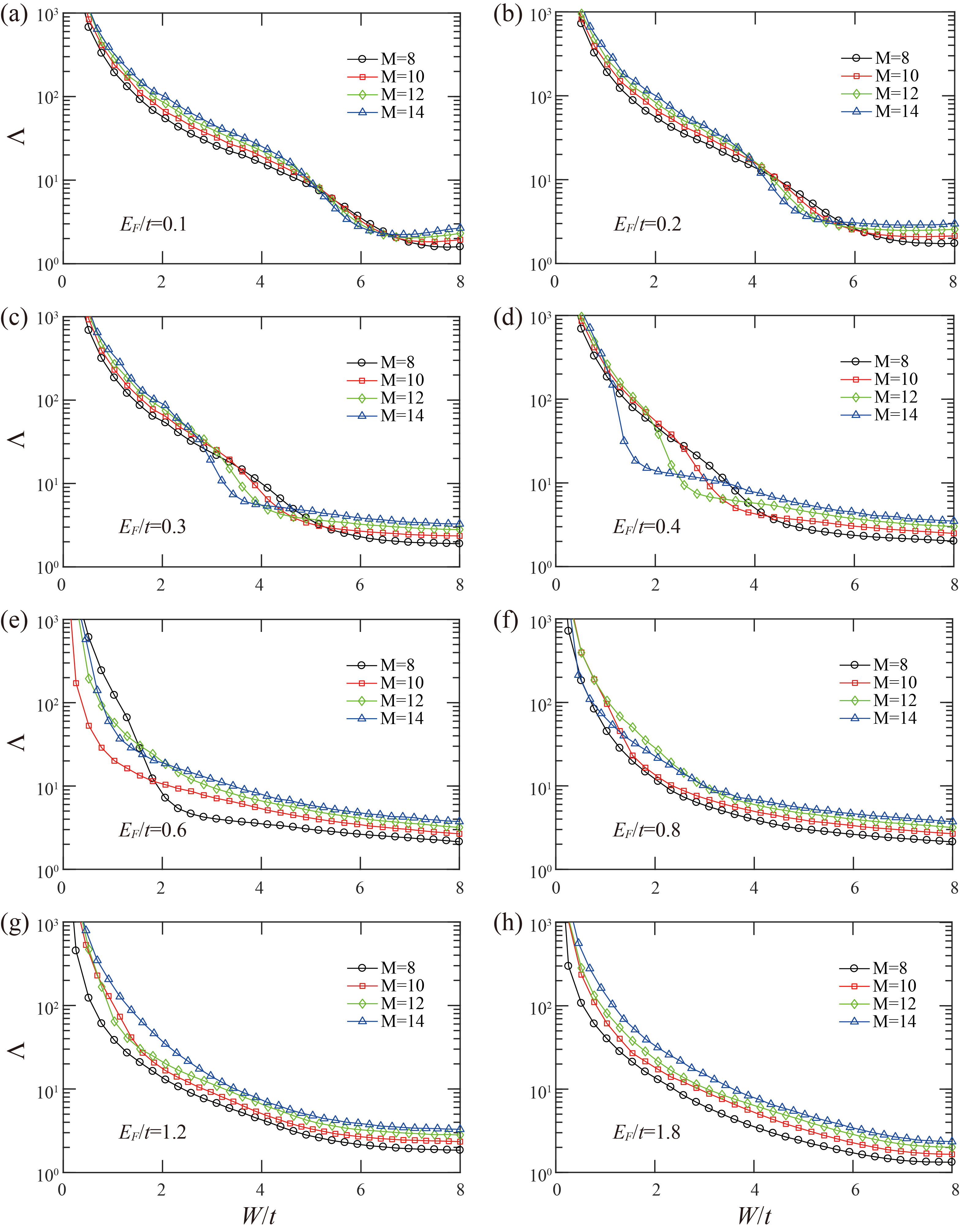}
  \end{center}
  \vspace{-0.2in}
\caption{\footnotesize{(a)-(h)}  Localization length as a function of 
$W/t$ for $E_F/t=0.1$, 0.2, 0.3, 0.4, 0.6, 0.8, 1.2, 1.8, respectively. 
$m_z=2.19m_0$ and $m_0=2.1t$  are fixed.}
  \label{s4}
  \vspace{-0.2in}
\end{figure}

To demonstrate how the intermediate CI phase change with the Fermi energy, 
we calculate the localization length for various $E_F$. In Fig. \ref{s3}, 
we plot the bulk energy bands projected onto the $k_z$-$E$ plane for the clean 
system with $m_z=2.19m_0$ and $m_0=2.1t$. The Fermi energies used for localization 
length calculations are denoted by the dashed lines for 
$E_F/t$=0.1, 0.2, 0.3, 0.4, 0.6, 0.8, 1.2, 1.8. The corresponding localization 
lengths are shown in Fig. \ref{s4}. Apparently, as $E_F$ increases from zero energy, the intermediate 
CI phase expands initially. This observation is consistent 
with the scattering analysis in Sec.~\ref{scattering}, since the internode scattering rate increases with $E_F$. Further increase of $E_F$, the linear dispersion relation fails  and the system becomes a conventional 3D metal when the Fermi energy deep inside the conduction band.

\begin{thebibliography}{99}
\bibitem{note1}
Strangely, the evidences of the transition resemble the conventional 
Anderson localization transitions at which the localization lengths of 
different sample sizes cross at the same point, and the uncorrelated 
on-site disorder is used in these studies so that internode scattering is 
comparable to intranode scattering and should be significantly important.
\bibitem{mz}Model parameter $m_z$ in the current work is 
defined as the $m_z+2m_0$ of Ref. \cite{Q10} to simplify the notation.
\bibitem{phase1}The conditions of the WSM phase of 
the model are: (1) $2m_0-t\leq m_z\leq 2m_0+t$ with 
the WNs located at $\bm{K}=(0,0,\pm
\cos^{-1}((m_z-2m_0)/t))$; (2) $-t-2m_0\leq m_z 
\leq t-2m_0$ with the WNs located at $
\bm{K}=(\pm\pi,\pm\pi,\pm\cos^{-1}(m_z/t
+2m_0/t))$; (3) $-t\leq m_z \leq t $ 
with the WNs located at $\bm{K}=(0,
\pm\pi,\pm\cos^{-1}(m_z/t))$ and 
$(\pm\pi,0,\pm\cos^{-1}(m_z/t))$.
The conditions of the CI phase of the model are: 
(1) $t<m_z<-t+2m_0$ with $\sum_{k_z}C(k_z)=M_z$; 
(2) $t-2m_0<m_z<-t$ with $\sum_{k_z}C(k_z)=-M_z$.
\end{thebibliography}
\end{document}